\documentclass[journal=jacsat,manuscript=article]{achemso}
\usepackage{graphicx}
\usepackage[version=3]{mhchem} 

\author{Arvind Arun Dev}
\email{arvind.dev@ipcms.unistra.fr}
\affiliation{Institut de Physique et Chimie des Matériaux de Strasbourg, UMR 7504 CNRS-UdS, 67034 Strasbourg, France}
\alsoaffiliation{Université de Strasbourg, CNRS, UMR 7140, 67000 Strasbourg, France}
\author{Peter Dunne}
\affiliation{Institut de Physique et Chimie des Matériaux de Strasbourg, UMR 7504 CNRS-UdS, 67034 Strasbourg, France}
\author{Thomas M. Hermans}
\affiliation{Université de Strasbourg, CNRS, UMR 7140, 67000 Strasbourg, France}
\author{Bernard Doudin}
\email{bernard.doudin@ipcms.unistra.fr}
\affiliation{Institut de Physique et Chimie des Matériaux de Strasbourg, UMR 7504 CNRS-UdS, 67034 Strasbourg, France}

\title[An \textsf{achemso} demo]
  {Fluid Drag Reduction by Magnetic Confinement}

\abbreviations{IR,NMR,UV}

\begin{document}



\begin{abstract}
  The frictional forces of a viscous liquid flow are a major energy loss issue and severely limit microfluidics practical use. Reducing this drag by more than a few tens of percent remain elusive. Here, we show how cylindrical liquid–in–liquid flow leads to drag reduction of 60–99\% for sub-mm and mm-sized channels, regardless of whether the viscosity of the transported liquid is larger or smaller than that of the confining one. In contrast to lubrication or sheath flow, we do not require a continuous flow of the confining lubricant, here made of a ferrofluid held in place by magnetic forces. In a laminar flow model with appropriate boundary conditions, we introduce a modified Reynolds number with a scaling that depends on geometrical factors and viscosity ratio of the two liquids. It explains our whole range of data and reveal the key design parameters for optimizing the drag reduction values. Our approach promises a new route for microfluidics designs with pressure gradient reduced by orders of magnitudes.
\end{abstract}

\section{Introduction}
Friction is a multifaceted problem existing in most physical processes and accounts for almost 25\% of energy loss in the world \cite{SayfidinovFrictionalEnergyLoss2018,HolmbergTribology2017}. Fluid friction or hydrodynamic viscous drag is decisive when designing energy efficient large scale flow systems  \cite{BrostowDRinFlow2008}. At smaller, microfluidic sizes, channels provide unique advantages for handling small volumes with reduced waste and manufacturing costs, ideal for drug delivery and discovery   \cite{DittrichMicrofluidicsDrugDiscovery2006, NeuziReviewDrugDiscovery2012}. Flow-focusing using microfluidic channels for drug delivery and chaotic mixing of reactants forms an essential part of pharmaceutical research  \cite{DamiatiDrugDelievery2018}. However, delivery of highly concentrated drugs through sub-mm diameter medical needles is difficult because of large drag due to the nonlinear increase in viscosity with increasing concentration \cite{SharmaViscosity2014}. This results in pumping forces beyond the range of manual injection \cite{JayaprakashLiquidInLiquid2020}. Furthermore, to limit shear damage, aggregation, and sedimentation in medically significant flows, like blood through tubes and arteries  \cite{ReinkeShearBlood1987}, reducing viscous drag is essential. Enabling drag control is also key in studying the response of cancer cells  \cite{MitchellCancerCellResponse2013} and viruses \cite{GreinShearStressVirus2019}. Reducing the hydrodynamic viscous drag and shear are key design issues and have led to many solutions, like mixing with additives \cite{LeeSpencerBiologicalAdditives2008}, surface chemical treatment \cite{WatanabeWaterRepellentWall1999}, or thermal creation of two-phase systems\cite{SaranadhiLeidenfrostSurface2016}; while Nature’s way, the Lotus effect\cite{EnsikatLotusEffect2011}, has steered research towards engineered (super)hydrophobic surfaces\cite{TuoAnisotropicSuperhydrophobicSurface2019,YoungSuperhydrophobicNanostructureDR2020,NeilSuperhydrophobicCoppertubes2009} with stabilized\cite{CarlSustainedSuperhydrophobicDesign2011,PinchasikCassieWenzelStabilization2016} liquid/gas interfaces  \cite{HuAirRings2017,ChoiLargeSlip2006,KaratayBubbleMattresses2013}. To overcome the drawback of the limited time stability of interstitial gas, oil/liquid infused surfaces have been proposed \cite{WongSlippery2011,SolomonLIS2014}.
Establishing liquid walls by means of an interstitial liquid lubricant is of particular interest for microfluidic applications, with several strategies documented in the literature\cite{HimaniSlipFlowLISMicrochannel2019,XuDynamicAirLiquidPocket2018,SamiraLubricationMicrofluidicChamber2011,WalshMicrofluidicswithfluidwalls2017,LiamHowSlipperyAreSlips2019,ScarattSlipLISwithAFM2020}. Nanostructured surfaces are shown to enhance lubricant retention thereby extending the lifetime of planar lubricated surfaces\cite{LaneyDelayedDepletionSLIPS_Nanostructure2021}. Hydrodynamic fluid focusing and microdroplets microfluidics are designed for surrounding the transported liquid material with an immiscible liquid  envelope\cite{RajawatCytometry2020,BaroudMicrodroplets2010}. Core annular flows, where the transported liquid is lubricated coaxially using a liquid of lower viscosity is proposed to handle highly viscous flows in microfluidics \cite{JayaprakashLiquidInLiquid2020}. However, forming a stable annular flow is difficult, and the risk of draining the lubricating liquid\cite{WexlerShearFaliure2015,ChapmanDepletionLubricant2021} limits the maximum achievable drag reduction.
The use of ferrofluids as confining liquid can improve the stability and draining issues, taking advantage of magnets to generate a force field that hold the confining ferrofluid in place. 
\begin{figure}[htp]
\centering
\includegraphics[width=0.9\textwidth]{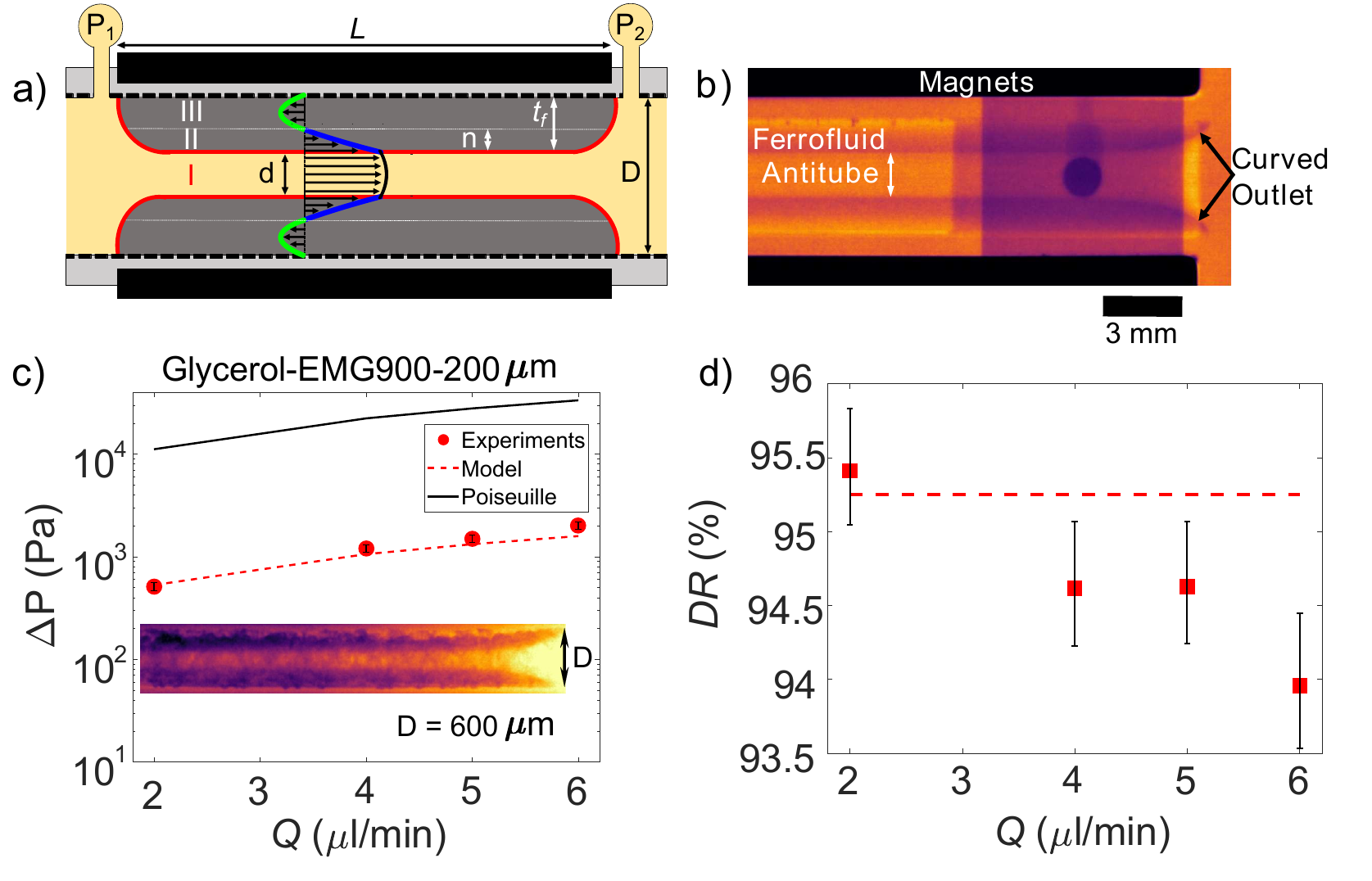}
\caption{\label{fig_1} a) Experimental setup for differential pressure measurement (P1, P2). An antitube of diameter $d$ inside a rigid cavity of diameter $D$ is shown with a velocity profile following the hypothesis of a 2-fluid, 3-region model. The red line between the ferrofluid and antitube depicts the liquid-liquid interface. b) X-ray absorption contrast image of the fluidic circuit, inside a cavity surrounded by magnets. The brighter central flowing liquid (antitube) is surrounded by a darker immiscible ferrofluid. c) Pressure-drop versus flow rate for an antitube of diameter 200 $\mu$m, with inset showing the X-ray image near the outlet (bright central part is antitube). Markers are experiments, the continuous line is the expectation from a Poiseuille flow of diameter $d$, and the dashed line is the prediction of the model detailed in the text below. d) Corresponding drag reduction with respect to Poiseuille flow. The viscosity ratio between the transported liquid and the ferrofluid is $\eta_r$=11.}
\end{figure}
Recent results using ferrofluid-infused surfaces showed superhydrophobic behaviour \cite{WangFerrofluidInfusedSurface2018}, and early pioneers \cite{MedvedevFlowSeparation1983} showed how pressure gradients can be reduced by an intermediate ferrofluid layer in large pipes\cite{MedvedevReducingResistance1987,KrakovHydrodynamicResistance1989}. Here we show how drag reduction, defined as the percentage change in friction factor\cite{WatanabeWaterRepellentWall1999,FukudaDragReduction2000,KaratayBubbleMattresses2013,Mirboddragreduction2017,Liudragreduction2020,Ayegbadragreduction2020}, is remarkably enhanced in milli- and micro-fluidic circuits using magnetic confinement of ferrofluid lubricants with appropriate magnet assemblies. Its key element is the implementation of our recent  design of quadrupolar confining magnetic field\cite{DunneLiquidflow2020}, capable of stabilizing the cylindrical flow of a diamagnetic liquid (called ‘antitube\cite{AntitubeMike}’) inside a ferrofluid envelope attracted towards the magnets. Our experimental system forms an ideal liquid–in–liquid tube system where the lubricating liquid forms a perfect concentric confinement for the flowing liquid, using a magnetic force design best suited to preserve the cylindrical geometry and providing optimum robustness of the system. This allows us to compare and verify a simple, axially-symmetric theory with experiments for a significant range of hydrodynamic parameters. The study goes beyond the assumption of unidirectional flow of the involved liquids, ideal for maximum drag reduction \cite{WalshMicrofluidicswithfluidwalls2017,SolomonLIS2014,JayaprakashLiquidInLiquid2020}. We previously discussed the expected and measured equilibrium diameter $d_{eq}$ of the antitube under static conditions (no flow), stabilized by the equilibrium between magnetic pressure and surface tension between the liquids\cite{DunneLiquidflow2020}. Here we investigate the dynamic case, where the fluid viscosities are key properties, and show how antitube circuits can exhibit drastically reduced pressure drop for significant ranges of viscosity and flow conditions.
\section{Experiments}
The fluidic cell design follows our previous work \cite{DunneLiquidflow2020} and is sketched in Figure 1a. 
To perform the flow experiments, the four-magnet assembly were housed in a 3D-printed support with built-in fluidic connectors for pressure measurements (cross-section Fig. 1a). For larger diameter antitubes ($>$ 1 mm) we used a support with a magnet spacing $w$ = 6 mm, internal cavity with diameter $D$ = 4.4 mm and inlet-outlet separation $L$ = 52 mm, and for sub-mm antitube $w$ = 0.9 mm, $D$ = 600 µm, and $L$ = 12 mm. The 3D printed cavity ($D$, Fig. 1a) is first filled with the non-magnetic liquid, then slowly replaced by injecting the ferrofluid with resulting formation of the antitube.\\
We control the diameter d, which is always $>$ $d_{eq}$, by varying the injected trapped volume of ferrofluid. We determined the antitube diameter from  X-ray absorption contrast images (see supporting information S.I. S1), either with 20 $\mu$m resolution setup\cite{DunneLiquidflow2020} (Fig. 1b), or a micro computed tomography system (RX-Solutions EasyTom 150/160) with 6 $\mu$m resolution, Fig. 1c inset. The antitube absorbs X-rays less than the ferrofluid, resulting in an image with a bright central region surrounded by two darker bands Fig. 1b). At the top and bottom of the image are the magnets which are opaque to X-rays at this thickness, and appear as two black regions, separated from the ferrofluid by the nearly-invisible 3D printed support. Near the ends of the magnets, the antitubes widen due to the fringe magnetic fields; fortuitously, these curved inlets and outlets improve the trapping of the ferrofluid (see S.I. S1). \\
We measured the pressure drop, $\Delta P$, between inlet and outlet, with Honeywell pressure transducers (HSCDLND001PG2A3, HSCDLNN400MGSA5) under a constant flow set by a syringe pump (Harvard apparatus PHD 2000). Fig. 1c shows the measured pressure drop as a function of flow rate of glycerol (Sigma Aldrich) flowing through EMG900 (FerroTech) ferrofluid (markers). This pressure drop is more than one order of magnitude lower than a Poiseuille flow in a solid channel with equivalent diameter (200 $\pm$ 6 $\mu$m). This illustrates how it is beneficial to have a cylindrical flowing liquid confined by a liquid interface rather than by solid walls.\\ 
The reduction in pressure drop is better described by the dimensionless friction factor, $f$. For a fluid with density $\rho$, the measured pressure drop $\Delta P$ resulting from a flow rate $Q$ through a tube of diameter $d$ and length $L$ is related to the experimental friction factor by \cite{KaratayBubbleMattresses2013}

\begin{equation}\label{Eqn_2.1}
    f_{exp}=\frac{\pi ^2 \Delta P d^5}{8 \rho Q^2L} 
\end{equation}
The drag reduction factor $DR$ is then defined as\cite{WatanabeWaterRepellentWall1999}
\begin{equation}\label{Eqn_2.2}
   DR= \frac{f_P-f_{exp}}{f_P} \times 100
\end{equation} 
\noindent which is the percentage change of the measured friction factor $f_{exp}$ compared to the factor $f_p$ under the same flow rate that follows Poiseuille’s law. At a solid wall boundary, $f_p$ = 64/Re , where the Reynolds number Re = $\rho$Ud/$\eta$ is defined for diameter $d$, density $\rho$ , viscosity  $\eta$ and average velocity $U$ corresponding to the flow rate $Q$. Due to the liquid-liquid interface, the zero-velocity boundary condition at the transported liquid wall (antitube–ferrofluid interface) does not apply, and the deviation from Poiseuille’s law results in hydrodynamic drag reduction. Fig. 1d) shows the experimental drag reduction (markers) based on Eq. 1 and Eq. 2 and from data in Fig. 1c), with values reaching 95.5 \%. 
\begin{figure}[htbp!]
\centering
\includegraphics[width=0.9\textwidth]{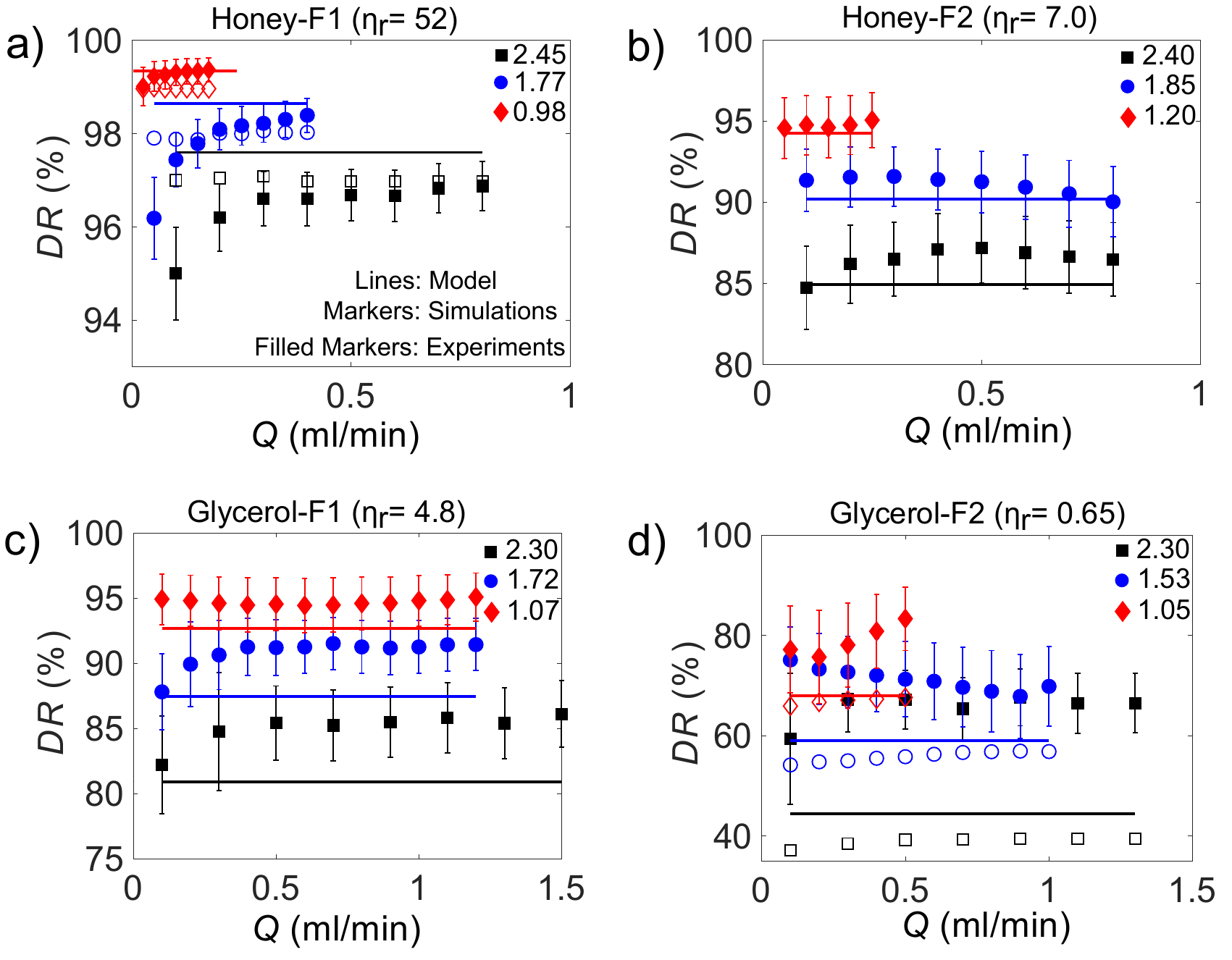}
\caption{\label{fig_2}  Drag reduction, $DR$, for three antitube diameters as a function of flow rate for a) honey as the transported liquid with APG314 ferrofluid (F1) as the confining liquid, b) honey with APGE32 ferrofluid (F2), c) glycerol with F1 and d) glycerol with F2. $\eta_r$ is the ratio between the antitube and ferrofluid viscosities. Legend shows the antitube diameter in mm. Lines, markers and filled markers compare model, simulations, and experiments, respectively.}
\end{figure}
A more complete insight into drag reduction possible values is gained by testing how it evolves under flow when varying the viscosity of both transported and confining liquids as well as the antitube diameter. We limit ourselves to viscous liquids, for measurable pressure differences and investigate near mm-sized channels for straightforward X-ray imaging.
The transported viscous liquids were glycerol (Sigma Aldrich) and honey (Famille Michaud), and the confining ferrofluids  used were APG314 (F1) and APGE32 (F2) (FerroTech), resulting in experiments on four combinations of magnetic and non–magnetic liquids.

\begin{figure}[htbp!]
\centering
\includegraphics[width=0.9\textwidth]{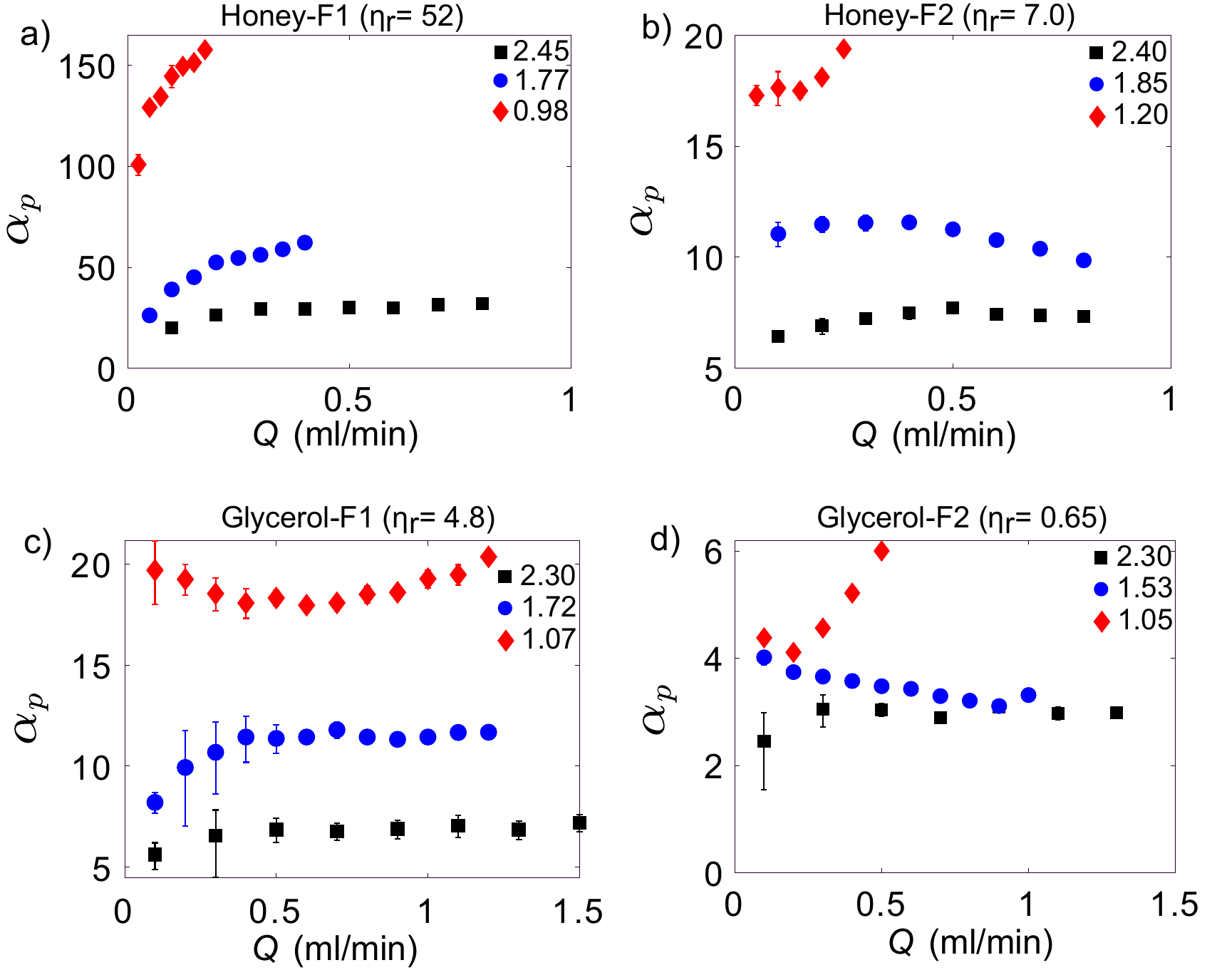}
\caption{\label{fig_3} Pressure drop reduction. a) Honey-APG314 ($\eta_r$=52), b) Honey-APGE32 ($\eta_r$=7.0), c) Glycerol-APG314 ($\eta_r$=4.8), d) Glycerol-APGE32 ($\eta_r$=0.65). $\alpha_p$ = $\Delta P_{exp}$ / $\Delta P_{p}$ is the ratio of pressure drop for solid wall tube to the antitube with identical diameter. The error bars are smaller than the size of markers where not visible.}
\end{figure}
The viscosities of these liquids were measured using a viscometer (Anton Paar MCR 502, see SI S2); for glycerol it is constant at 1.1 Pa.s and for honey 11.99 Pa.s under strain rate of 100 s$^{\text{-}1}$ (maximum strain rate in our experiments with honey in antitube is below 10 s$^{\text{-}1}$). Our measurements span a viscosity ratio $\eta_r$ between transported and confining liquids both larger ($\leq$52) and smaller ($\geq$0.65) than one, with data for three different antitube diameters and four different viscosity ratios summarized in Fig. 2 and Fig.3.\\
All 4 combinations of transported and confined liquids show remarkably high drag reductions ranging from 60 \% to 99.3 \% (Fig. 2).  Drag reduction increases with increasing viscosity ratio, with a maximum for Honey-F1 with $\eta_r$ = 52 (Fig. 2a). On the contrary, no drag reduction is expected when the ratio goes to zero, describing an infinitely viscous envelope, or a solid wall. However, in contrast to prior expectations \cite{SolomonLIS2014}, large drag reduction can still be achieved, even if the confining ferrofluid has a larger viscosity than the transported one, such as for Glycerol-F2, where a drag reduction of up to 80\% is observed (Fig. 2d). Additionally, the drag reduction increases with decreasing antitube diameter (Fig. 2a,b,c,d), which is beneficial when miniaturizing the fluidic circuit.\\
Alternatively, the large drag reduction can be expressed as an improvement ratio $\alpha_p$ = $\Delta P_{exp}$/$\Delta P_{p}$,  which compares the measured pressure drop in a liquid-walled interface, $\Delta P_{exp}$, to a solid-wall interface, $\Delta P_{p}$, of equivalent diameter, d. Fig. 3 shows more than two orders of magnitude of improvement ($\alpha_p$) can be achieved. 
For $\eta_r$  = 52, the antitube system results in 157 times less pressure drop than the solid walled tube (Fig. 3a, red markers). Interestingly a viscosity ratio $\eta_r$ = 0.65 still results in an almost 6 times smaller pressure drop (Fig. 3d), and the improvement ratio $\alpha_p$ increases with decreasing antitube diameter (Fig. 3a,b,c,d). 
\section{Modelling}
We explain our results using a two-fluid model, following the previous pioneering works \cite{KrakovHydrodynamicResistance1989}, based on the steady-state one dimensional Navier-Stokes equation with velocity as a function of radius in a cylindrical geometry, $u$ = $u$($r$), under modified boundary conditions. Note that we present below equations that differ from those presented in literature\cite{KrakovHydrodynamicResistance1989}, motivated by the need to compare the analytical expressions to numerical simulations. We checked that the outcomes of both analytical approaches are identical, under the hypothesis of non-deformable interfaces detailed below. A key ingredient of the model is the occurrence of a counter flow within the confining ferrofluid (Fig. 1a) resulting from avoiding drainage of the ferrofluid by means of the magnetic sources. This suppression is due to the non-uniform magnetic fields at the inlet and outlet opposing any egress of ferrofluid. As the ferrofluid cannot escape but noting that a) flux must be conserved and b) that the drag reduction should result from a non-zero velocity at the ferrofluid-antitube interface, a return path for the ferrofluid flow must exist. The simplest hypothesis is illustrated by the velocity profile in Fig. 1a, where we define three regions: I inside the antitube, II the part of the ferrofluid that travels alongside the antitube flow, and III where counter-flow occurs. The non-dimensional governing equations for the three regions (i = I, II, II) are given by
\begin{equation} \label{Eqn_3.1}
  \frac{1}{Re_{i} \thinspace r^\star}  \frac{\partial}{\partial r^\star}\left(r^\star \frac{\partial u^\star_{i}}{\partial r^\star}\right)=\frac{\partial P^\star_{i}}{\partial z^\star}
\end{equation}
where $u^\star_{i}$,$r^\star$  are dimensionless velocity and coordinates scaled by the average velocity $u_m$  and the diameter $d$ of the region I (antitube), respectively. The dimensionless pressure is defined as $P^\star_{i}$=$P_i$⁄$\rho_i u_m^2$  with the corresponding Reynolds numbers for each region being  $Re_i$=$\rho_i u_m d$⁄$\eta_i$ . Note that the pressure gradients along the main flow in both ferrofluid regions are equal, under the hypothesis that these two regions do not mix, resulting in the absence of pressure gradient along $r$, and therefore  $\frac{\partial P_{II}^\star}{\partial z^\star}=\frac{\partial P_{III}^\star}{\partial z^\star}$. \\

\noindent The magnetic-nonmagnetic interface is modelled as a non-deforming fixed liquid wall. This hypothesis limits us to experiments under low-enough flow values. The case of diameter depending on flow and position along $z$ is beyond the scope and minimal set of hypotheses of this paper, as it is related to the interface stability of non-miscible liquids, and the need to introduce explicitly magnetic stress forces in the model. We checked by X-ray imaging that we are not in this case, aware the deviations for the model will increase with the flow rate, as deformations cannot be neglected anymore. \\
We assume here that the pressure gradients in regions I and II are different, to ensure a non-deformable interface. Note that earlier research\cite{KrakovHydrodynamicResistance1989} presented the analytical model as a 2 fluid model with equal pressure gradient, $\frac{\partial P_{I}^\star}{\partial z^\star}=\frac{\partial P_{II}^\star}{\partial z^\star}$., but since the numerical simulations do not take into account the magnetic forces field, it is essential to define the magnetic-nonmagnetic interface as non-deformable wall resulting in  $\frac{\partial P_{I}^\star}{\partial z^\star} \ne \frac{\partial P_{II}^\star}{\partial z^\star}$. (see section 4 and SI S3). We insist, that so long as the shear stress and velocity boundary conditions at the magnetic-nonmagnetic interface are satisfied, the assumption of pressure gradients do not affect the drag reduction calculations.\\
Our hypothesis allows us to consider a diameter $d$, set experimentally by the amount of ferrofluid trapped in the cavity, independent of the flow rate and treat the problem numerically to match our experimental conditions. These governing equations are solved for all velocities ($u^\star_{I}$,$u^\star_{II}$,$u^\star_{III}$), pressure gradient $\frac{\partial P_{III}^\star}{\partial z^\star}$ and thickness $n$ of region II, using boundary conditions depicted in the Fig. 1b:  finite velocity at the antitube centre, zero velocity at the solid wall and at interface of II and III, continuity of velocity and shear stress at interfaces I-II and II-III. Along with these boundary conditions, the volume conservation of ferrofluid dictates that the flow rate in region II and III must be equal, $Q_{II}^\star=Q_{III}^\star$. We present two models, one with no assumption (full model) and another with assumption $\frac{\partial P_{II}^\star}{\partial z^\star} = 0$, which explicitly shows the contribution of geometric and fluid parameters for drag reduction.
In the full model Eq. 3 expands to
\begin{equation} \label{Eqn_3.1_RegIII}
  \frac{1}{Re_{III} \thinspace r^\star}  \frac{\partial}{\partial r^\star}\left(r^\star \frac{\partial u^\star_{III}}{\partial r^\star}\right)=\frac{\partial P^\star_{III}}{\partial z^\star}
\end{equation}
\begin{equation} \label{Eqn_3.1_RegII}
  \frac{1}{Re_{II} \thinspace r^\star}  \frac{\partial}{\partial r^\star}\left(r^\star \frac{\partial u^\star_{II}}{\partial r^\star}\right)=\frac{\partial P^\star_{II}}{\partial z^\star}
\end{equation}
\begin{equation} \label{Eqn_3.1_RegI}
  \frac{1}{Re_{I} \thinspace r^\star}  \frac{\partial}{\partial r^\star}\left(r^\star \frac{\partial u^\star_{I}}{\partial r^\star}\right)=\frac{\partial P^\star_{I}}{\partial z^\star}
\end{equation}
Solving Eq. 4 to Eq. 6 with the boundary conditions mentioned gives the analytical expressions of the flow rate in the antitube as:
\begin{equation}\label{FlowAntitube}
   Q^\star_I= \frac{\pi Re_I}{4}\frac{\partial P^\star_I}{\partial z^\star}\left[\frac{1}{32}+a_3-a_4-a_5\right]
\end{equation}
and the resulting friction factor written as \begin{equation}\label{FrictionFactor}
 f_A=64/Re_I \thinspace \beta   
\end{equation}
where
\begin{equation}\label{Beta}
 \beta=32(a_5+a_4-a_3)-1 
\end{equation}
and $a_3$, $a_4$, $a_5$ are scalar constants that can be expressed as explicit functions of $d$, the thicknesses $n$ of the region II and $t_f$ of the ferrofluid.\\ 
This ‘full model’ is therefore fully analytically solvable; however, Eq. 8 and Eq. 9 together presents a complex expression where the contribution of fluid and geometric properties are hidden. Simplified expressions illustrating better the key contributions to drag reduction are obtained by neglecting the pressure gradient in region II. We show below that this artificial hypothesis has limited impact on the accuracy of the results. This approximation $\frac{\partial P_{II}^\star}{\partial z^\star} = 0$ results in a simplified expression for $\beta$:

\begin{equation}\label{BetaApproax}
 \beta_0=1+4\ln\bigg(1+\frac{2n_0}{d}\bigg)\eta_r 
\end{equation}
where $n_0$ is the thickness of the ferrofluid in region II under this approximation. $\eta_r=\eta_a/\eta_f$ is the ratio of viscosity of antitube liquid ($\eta_a$) and ferrofluid ($\eta_f$). Note that we take the viscosity of the ferrofluid at saturation in a magnetic field (see SI S2). Complete and explicit equations for the scalar constants and simplified model are given SI 4.

\subsection{4.	Numerical visualization of counter flow}
Greater confidence in our equations results were gained by computational fluid dynamics (CFD) simulations using ANSYS CFX 18, where detailed information on the occurrence of counter flow and the resulting velocity vector field were obtained.
\begin{figure}[htbp]
\centering
\includegraphics[width=0.9\textwidth]{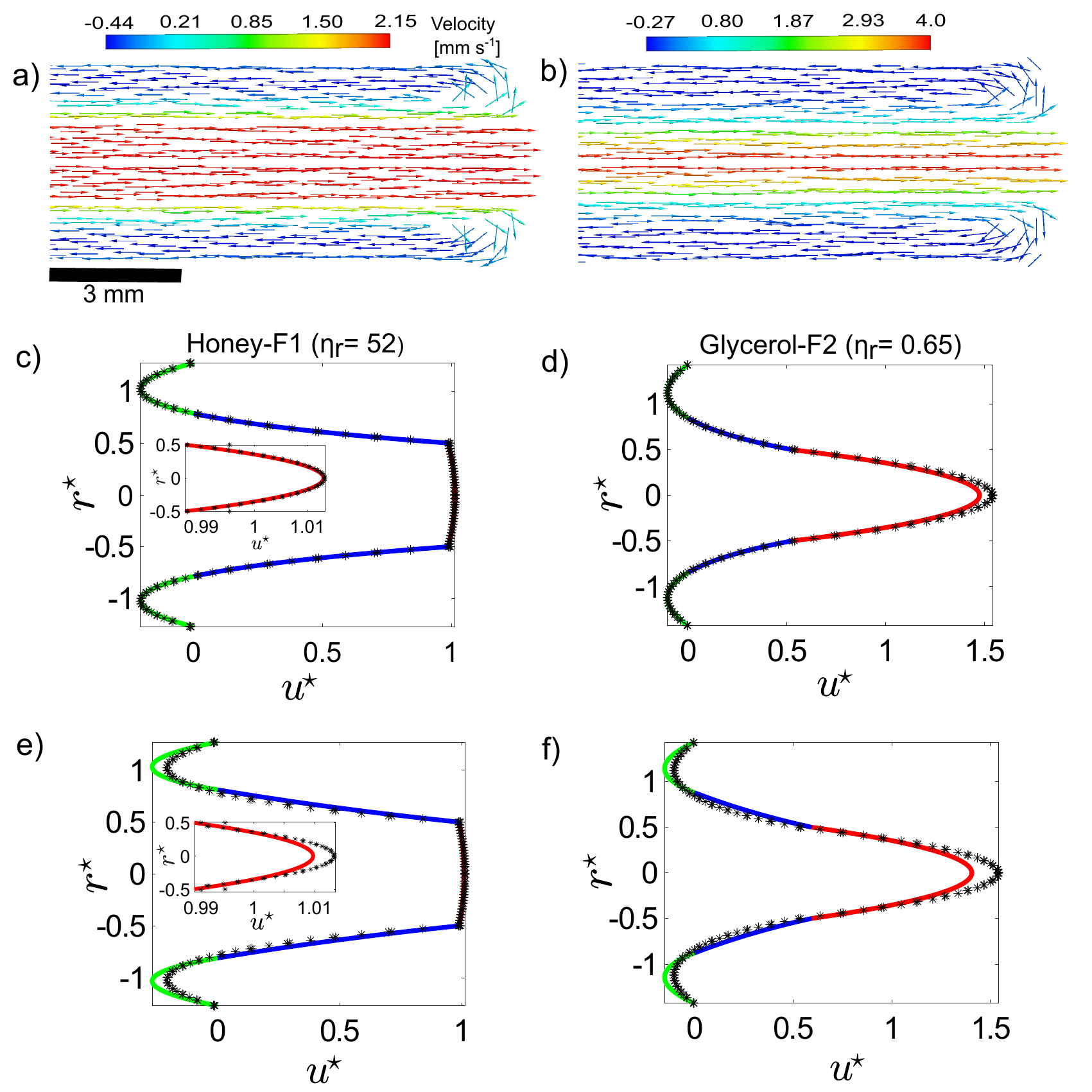}
\caption{\label{fig_4} . Simulated visualization of counter flow. a) and b) Velocity vectors near the outlet showing counter flow for Honey-F1 and Glycerol-F2 respectively. c) and d) comparison of the non-dimensional velocity profile of the full analytical model and simulations in Fig. 4(a) and Fig. 4(b). e) and f) comparison using simplified analytical model neglecting pressure gradient term. Lines are model predictions and markers are simulations calculations. Inset shows the magnified view of velocity profile in an antitube. The antitube diameter for left and right column is 1.73 mm and 1.54 mm respectively. $\eta_r$ is viscosity ratio.}
\end{figure} 
In the numerical simulations, we solve the three-dimensional Navier–Stokes equation by considering the magnetic-nonmagnetic interface as a non-deformable fixed liquid wall of infinitesimal width. The axial velocity and shear stress are equal on both sides of the wall, i.e., continuous across the interface. Since the magnetic field gradient only exists in the radial direction far from the inlet or outlet, no magnetic body force is considered on the ferrofluid. The finite curved edges of ferrofluid near inlet and exit of the flow are modelled as a free slip wall with zero normal velocity (no flow across the curved edges) (See SI S3 for numerical algorithm).  The numerical simulations also consider the shear dependent viscosity of ferrofluids. As illustrated in Fig. 4, a counter flow occurs in the ferrofluid close to the outer wall for Honey-F1 with $d$ = 1.73 mm (Fig. 4a) and Glycerol-F2 with $d$ = 1.54 mm (Fig. 4b), and flow rate fixed at $Q$ = 300 $\mu$l min$^{\text{-}1}$. Good agreement between numerical and analytical velocity profiles using the full model for both Honey-F1, $\eta_r$ = 52 (Fig. 4c), and Glycerol-F2,$\eta_r$ = 0.65, (Fig. 4d) validates our numerical algorithm. To more accurately model the experimental system shown in Fig. 1, we extended the simulations to consider the finite length of a device, and the effect of fringe fields on the shape of interface at the inlet and outlet (curved inlet and outlet, Fig. 1b, see SI S3), beyond the hypothesis of the infinite tube of the analytical model.
Numerical simulations were found to reproduce well the drag reduction data in Fig. 2a), while systematically underestimating observed drag reduction for $\eta_r$ = 0.65 (Fig. 2d). Note that the numerical drag reduction is calculated using Eq. 1 and Eq. 2 with $\Delta P$ obtained from numerical simulations.\\
\noindent The simplified model obtained by neglecting pressure gradient term in region II is also compared with the numerical simulations shown in Fig. 4e and Fig. 4f. The difference between the two analytical models is apparent on comparing Fig. 4c with Fig. 4e and Fig. 4d with Fig. 4f.
\subsection{Discussion}
The analytical model with a minimum set of hypotheses  presented here captures reasonably well the experimental measurements for all measurements presented in Fig. 1 and Fig. 2 for mm and sub-mm channels. The modified Reynolds number is the non-dimensional governing parameter and spans four orders of magnitude in our experiments. \begin{figure}[htbp!]
  \centering
  \includegraphics[width=0.7\textwidth]{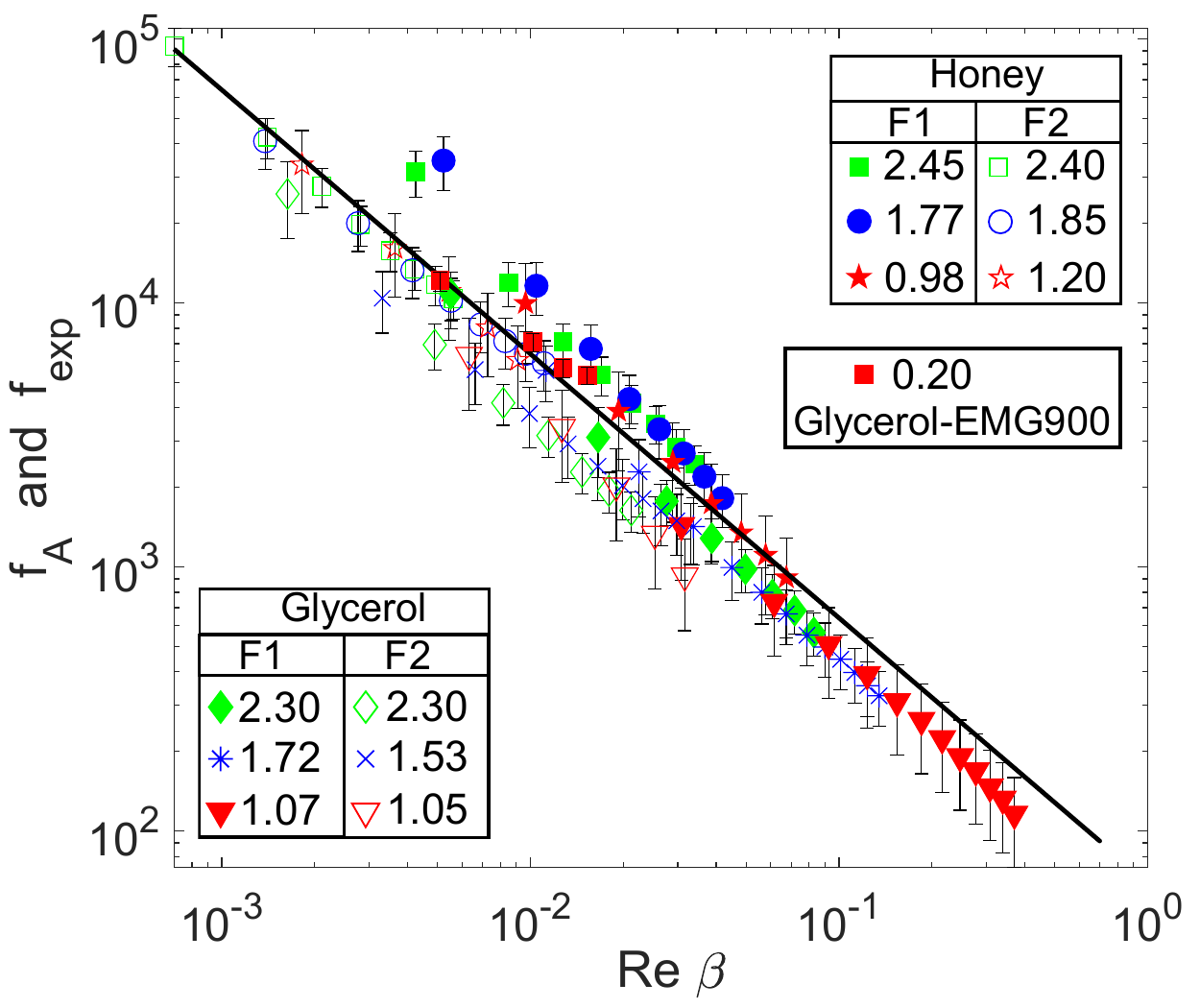}
  \caption{\label{fig_5} Comparison of experimental fexp (markers) and analytical $f_A$ (line) friction factors. Inset gives the diameter of antitubes (mm) for each case. Re denotes the Reynolds number and $\beta$  is the scaling factor (= 1 for solid-walled tube).}
  \end{figure}
  For a cylindrical tube flow $f$ = 64/Re , the liquid-in-liquid system results in $f_A$=64/Re$\beta$ for the same flow rate $Q$, illustrating how the friction factor is reduced by the scaling factor $\beta$. Fig. 5 illustrates how the calculated friction factor $f_A$ which is a function of the modified Reynolds number (from Eq. 3.6) compares with the experimental one $f_{exp}$ computed using Eq. 2.1. The friction factor $f_A$ is like the fit parameter presented in literature\cite{KrakovHydrodynamicResistance1989}. We provide here full analytical expressions for $f_A$, necessary for the equation of the drag reduction defined as percentage change in friction factor $f_A$ with respect to friction factor for Poiseuille flow $f_p$. Similar rescaling of the friction factor was also discussed for a highly water-repellent wall system\cite{WatanabeWaterRepellentWall1999}. Although the model cannot completely account for minor offsets observed at low viscosity ratio values, where deformation of the antitube starts to take place, Fig. 5 nevertheless illustrates the broad range of fluidic conditions that the model can apply to. Note that a significant variation of drag reduction with flow rate is found for the largest viscosity ratio ($\eta_r$ = 52).
In such cases, the pressure drop becomes very small at small flow rates, and the difficulties in neglecting the pressure loss related to the interconnects and pressure indicators limit the reliability of the data, systematically underestimating the drag reduction values. We also expect that more complicated fluid velocity profiles along $z$ can develop, especially for high viscosity confining liquids where the magnetic/non-magnetic interface might deform significantly due to the high pressure drop but taking them into account is beyond our current model.
 \begingroup
\begin{table}\label{table_1}
  \begin{center}
\def~{\hphantom{0}}
  \begin{tabular}{lcccccc}
  \hline
      $Case$ & d (mm) & $t_f$ (mm) & $n$ (mm) & $\beta$ & $n_0$ (mm) &  $\beta_0$\\[4pt]
      \hline
         & 2.45 & 0.98 & 0.34 &	41.6 &	0.38 & 58.68\\
 Honey-APG314 & 1.77 & 1.32 & 0.49 &	74 & 0.53 &	98.3\\
    & 0.98 & 1.71 & 0.67 & 151.68 &	0.73 &	183.35\\   
 \hline
         & 2.40 & 1.00 & 0.36 &	6.74 &	0.39 & 9.12\\
      Honey-APGE32   & 1.85 & 1.28 & 0.47 & 10.21 & 0.51 & 13.39\\
    & 1.20 & 1.60 & 0.61 & 17.44 & 0.67 & 21.49\\
  \hline
         & 2.30 & 1.05 & 0.38 &	5.24 &	0.41 &	6.95\\
 Glycerol-APG314  & 1.72 & 1.34 & 0.50 & 7.98 &	0.54 &	10.26\\
  & 1.07 & 1.67 & 0.65 & 13.62 & 0.70 & 16.46\\
 \hline
 & 2.3 & 1.05 & 0.38 & 	1.57 & 0.41 &	1.8\\
 Glycerol-APGE32  & 1.53 & 1.44 & 0.54 & 2.11 & 0.58 & 2.44\\
  & 1.05 & 1.68 & 0.65 & 2.74 & 0.71 & 3.12\\
 \hline
  \end{tabular}
  \caption{Comparison of the full and simplified models}
  \label{tab:kd}
  \end{center}
\end{table}
\endgroup
\noindent The simplified expression $\beta_0$ in Eq. 10, illustrates the deviation from the asymptotic $\beta_0$ = 1 value for solid walls, and is a simple but explicit way to quantify the reduction in the friction resulting from liquid-in-liquid flow. Compared to $\beta$, this approximation underestimates the frictional drag (visible in Fig. 4) and becomes more apparent when the thickness of the ferrofluid decreases (see Table 1). However, $\beta_0$, explicitly reveals the contributions of the antitube geometry and fluid properties: the drag reduction can be tuned by the choice of viscosities ($\eta_r$) and the amount of ferrofluid trapped in the device cavity, i.e. antitube diameter $d$, and ferrofluid thickness $t_f$.

\subsubsection{Conclusions}
We have studied the flow of viscous liquids through cylindrical liquid-in-liquid tubes where the confining liquid is held in place by a quadrupolar magnetic field. Our results show that drag reductions exceeding 99\% can be achieved by exploiting the non-zero velocity of a viscous liquid at its interface with the encapsulating liquid. The friction reduction is quantified by rescaling of the Reynolds number with a factor $\beta$ in Eq. 9. The drag reduction improves when decreasing the diameter of an antitube relative to its surrounding ferrofluid, or when increasing the ratio of the antitube to ferrofluid viscosities. The former is relevant for the needs and length-scales of microfluidics, while the latter indicates that large drag reduction is expected when flowing highly viscous liquids. Moreover, with antitube diameters as small as 10 $\mu$m already achievable\cite{DunneLiquidflow2020}, antitube diameter to ferrofluid thickness ratios of order 100 are within reach. Therefore, very large drag reduction in microfluidic channels is possible for a broad range of confining liquid viscosities. Downsizing or designing magnetic force gradients along the flow direction can also further enhance the stability of the ferrofluid against shearing, paving the way to both high velocity and low viscosity fluidic applications in domains ranging from nanofluidics to marine or hydrocarbon cargo transport.

\begin{acknowledgement}

This project has received funding from the European Union’s Horizon 2020 research and innovation programme under the Marie Skłodowska-Curie grant agreement No 766007. We also acknowledge the support of the University of Strasbourg Institute for Advanced Studies (USIAS) Fellowship. This work was also partly supported by IdEx Unistra (ANR 10 IDEX 0002), and by SFRI STRAT’US project (ANR 20 SFRI 0012) and EUR (QMat-ANR-18-EUR-0016) under the framework of the French Investments for the Future Program. We thank A. Zaben, Prof. Cēbers and Dr. Kitenbergs of MMML lab, Riga, Latvia for ANSYS CFD software facility and magnetoviscosity measurements. The authors thank Prof. Gauthier, Antoine Egele and Damien Favier of Institut Charles Sadron, Strasbourg, France for the use of the EasyTom X-Ray Tomography facility. 

\end{acknowledgement}

\begin{suppinfo}
The supporting information consist of a PDF document with 
\begin{itemize}
    \item Procedures for characterising antitube diameters.
    \item Viscosity measurements of antitube liquids and confining ferrofluids.
    \item Numerical simulation procedures and flow chart.
     \item Details of analytical model used in this article.
\end{itemize}

\end{suppinfo}

\bibliography{FluidDR_Dev}

\end{document}